# RefDB: The Reference Database for CMS Monte Carlo Production

Véronique Lefébure
*CERN (Geneva) & HIP (Helsinki Institute of Physics)*
Julia Andreeva
*CERN (Geneva) & UC Riverside*

RefDB is the CMS Monte Carlo Reference Database. It is used for recording and managing all details of physics simulation, reconstruction and analysis requests, for coordinating task assignments to world-wide distributed Regional Centers, Grid-enabled or not, and trace their progress rate. RefDB is also the central database that the workflow-planner contacts in order to get task instructions. It is automatically and asynchronously updated with book-keeping run summaries. Finally it is the end-user interface to data catalogues.

## 1. INTRODUCTION

RefDB is the CMS [1] Monte Carlo Reference Database. It was designed and implemented during the last two months of 2001 and used for the first time during the Spring 2002 production in which 20TB of data was produced for the DAQ TDR [2]. RefDB is also now used by the CMS Grid prototypes in both the USA and Europe to allow them to produce data of use to the CMS physics community. RefDB will be used for the CMS Computing Data Challenge and pre-challenge to start in July 2003.

RefDB has four functionalities:

1. It serves as a production-request submission system for the physicists. There are currently more than 20 persons registered as Requestors, with more than 2000 requests and 300 parameter files recorded. Web-forms allow physicists to define all the request details, including every parameter of physics software such as PYTHIA [3]. Previously-registered sets of parameters can also be used. Data that has not yet been produced can be selected as input for new requests, thereby satisfying some of the CMS requirements for a Virtual Data Catalogue. Production tasks cover generation of primary particles, simulation of detector response at different beam luminosities, filtering, cloning, extension of data, reconstruction of tracks and physics object creation of final analysis data such as ntuples or ROOT [4] files.

2. RefDB is accessed via a web-server by the workflow-planner tool (e.g. McRunjob [5]) as the central source of production instructions. This mechanism allows a high level of automation in the production process. Chaining of production steps is also supported.

3. RefDB is used for the coordination and monitoring of the world-wide distributed production. More than 20 sites participated in the Spring 2002 production and Grid prototypes have joined since. Jobs are monitored locally (e.g. using BOSS [6]) and metadata is transferred asynchronously to RefDB for validation and book-keeping. Quasi-online statistics and production rate are thus available, providing an essential overall view of the progress of the production.

4. Finally, RefDB is a metadata catalogue for the use of the physicists who want to find out what the available data is and how to access it. Physicists may also just want to have access to parameter files for private work.

MySQL is used as RDBMS. The database is hosted at CERN and is accessed exclusively via a web-server using Php scripts and *.htaccess* security system. The current size of the database amounts to 24 MB.

The general data flow presented on Figure 1 is the following: a pre-registered physicist submits a computing request. The request is assigned by the production coordinator to a regional centre by communicating to the production operator an AssignmentID that is used as input argument to the workflow-planner tool. The workflow-planner contacts RefDB (with *wget*) to get the information needed to decompose the task into atomic tasks and to create the job scripts. Successful jobs send summary files by e-mail, which are used for validation and book-keeping.

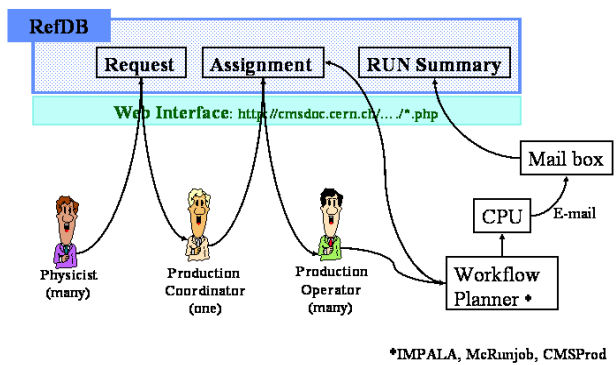

**Figure 1: General Data Flow in RefDB**

## 2. PHYSICS PRODUCTION REQUESTS

The definition of an atomic production request (also called *Derivation* in the Virtual Data Language) consists in specifying the following four fields, which are described in details below: the executable to be run (also called *Transformation*), the input physics parameters, the input data and number of events, and the input production parameters. Uniqueness of requests is checked. When a request is submitted, the production coordinator is automatically notified by e-mail, as well as the physics group coordinator and the requestor.

## 2.1. Executable

The executable is selected by the physicist according to the software name and version, and the executable name, all are pre-registered in RefDB. For example: "ORCA ORCA_7_1_1 writeAllDigis".

Binaries are distributed to the production sites using the DAR tool [7]. Physics code is versioned and managed using CVS [8] and SCRAM [9] but use of private user code is supported by a loading system.

Executable names are recorded in RefDB together with the log file parsing codes and schema needed for job monitoring, such as the ones used by BOSS (see Figures 2 and 3). The post-process parser is used to create the summary file that is sent automatically to the CERN mailbox by successful jobs. Validation of the summary files and update of RefDB occurs asynchronously and regularly.

## 2.2. Physics Parameters

Input physics parameter files are made of one or more file fragments that are registered into RefDB. The modularity of the parameter files allows an easy re-use of some coherent set of the parameters for different requests. For example, detector response parameters are separated from beam-luminosity-dependent parameters.

The physicist creates a file using a web-form for selecting the key=value pairs (s)he needs. If the file that is being created actually already exists, its ID is returned to the user for re-use. Each key=value pair is stored in a map table and files are recorded in RefDB as a string of key=value pair ID's (see Figure 4). Specialized scripts are responsible for formatting the actual ASCII files. These physical files are stored under CVS, and retrieved by the workflow-planner when jobs are created.

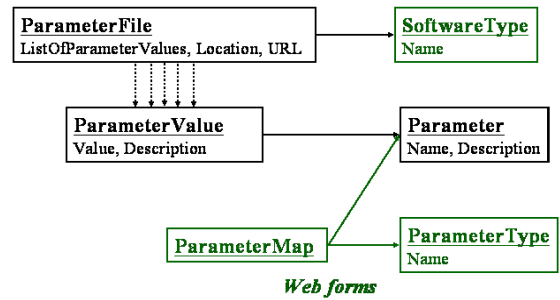

Figure 4: Parameter tables in RefDB

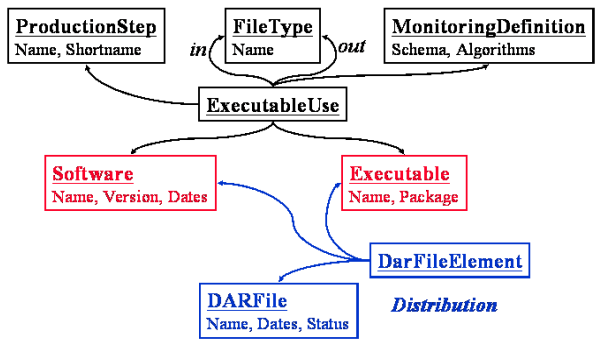

**Figure 2: Software and Executable tables in RefDB**

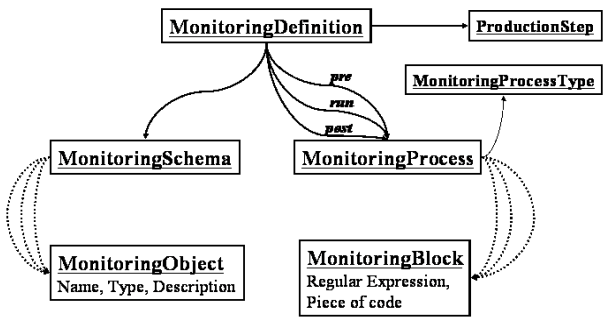

**Figure 3 Monitoring tables in RefDB**

## 2.3. Input data

The user has to specify the number of events to be produced, and can select, if needed, the logical name of the input collection of data to be processed. When a new dataset is created, the user has to specify its name. Data that has not yet been produced can nevertheless be selected as input data for a new request. It is checked that the type of data of the selected collection can indeed be handled by the selected executable. That information about compatibility between executable and input data is stored in RefDB in the ExecurtableUse table (see Figure 2).

There is a one-to-many relationship between a Dataset and a Collection. Datasets are uniquely defined by the physics channel that is simulated, and by the detector configuration (geometry, material, magnetic field specifications). For each Dataset new Collections are created each time a production cycle is requested. Production cycles are defined by the executable and the parameters to be used (see Figure 5).

**TUKT002**

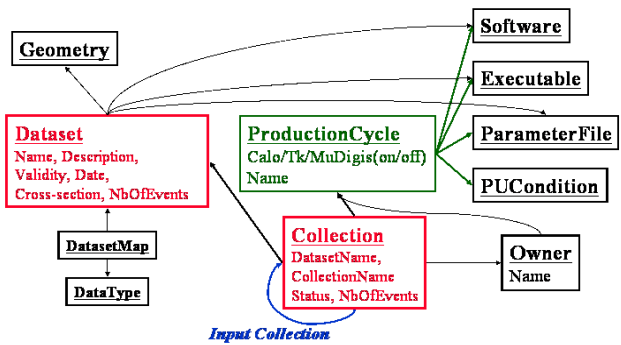

**Figure 5: Dataset and Collection tables in RefDB**

## 2.4. Production parameters

Production parameters are hidden from the physicists, and defined by the production coordinator. They specify how to cluster and name the output data, how to monitor the executable, how to split a large request into smaller jobs, including consequently the determination of variables that have to be changed for each job, such as run numbers and random number seeds.

## 3. PRODUCTION ASSIGNMENTS

Requests or slices of requests are assigned to regional centers by the production coordinator using a web-interface (see Figure 6). The regional center is notified, both by e-mail and from the web-page, of the new assignments. The assignment of requests is not an automated task because one has to take into account many time-dependent criteria such as production farm and operator availability and performance. Also local physics interest is taken into account as much as possible in order to minimize file transfers. Finally, requests are assigned according to the priority that is specified by the requestor at submission time. When regional centre consists of many farms, the local production coordinator is free to re-assign the tasks from the regional centre to the production site, using a web-tool. In that way, statistics about production rates per farm is still recorded centrally. Since assignment status is updated quasi online, it is easy to monitor the local and global production rate, and to estimate the time to completion.

### 3.1. Production Instructions

The Assignment ID's are the key used by the workflow-planner to retrieve all production instructions. The production instructions are listed here. They are invariant of the production steps:
- Executable Name, Software and Version
- Parameter file URL
- Job splitting instruction file URL
- Monitoring file URL

**TUKT002**

- Logical file name of configuration data (such as geometry file)
- Dataset name, Production Cycle (for information only)

The parameter file is the concatenation of all file fragments defined by the user, with placeholder strings for variables that are job dependent. The job splitting instruction file is a table containing, in the header line, the list of placeholder strings, and each following line gives the values by which they have to be replaced in the parameter file. The monitoring file contains the schema and the parsers needed to produce the run summary file that gives the information needed for book-keeping.

Chain assignments also exist: they point to an ordered list of Assignment ID's and are used for running sequentially several executables in a single batch job. Chain assignments may reduce considerably the production operations when only the final data product is needed and the intermediate files may be destroyed.

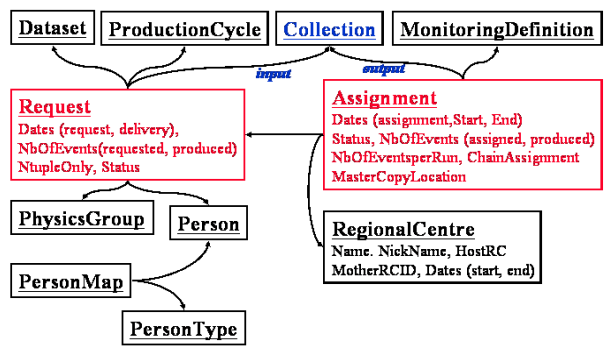

**Figure 6: Request and Assignment tables in RefDB**

## 4. PRODUCTION BOOK-KEEPING

For book-keeping of run-related information, one table per dataset is created, with one row per generation run. For each production cycle the following is recorded:
- Run number
- Random number seeds
- Cross-section
- Logical output file name
- Status
- Assignment ID
- Number of input events
- Number of output events
- Execution duration
- Size of output data

## 5. DATA CATALOGUE

The data catalogue implemented in RefDB is a catalogue of catalogues. It is a simple system of maps that store associations between Collections and external catalogues such as Objectivity/DB [10] or POOL [11] catalogues, mass storage system or disks, all being registered in a table together with their physical location. Scripts are run regularly in order to check the completeness of the catalogues: all files relative to a given collection of data have to be registered in a catalogue. A dataset query web-interface allows physicists to look for a set of datasets according to criteria such as name, physics channel, software,..., and to find out in which catalogue the data is published.

## 6. DEVELOPMENTS AND PROSPECTS

Several small developments are still in progress as we understand better and better the system, and backward compatibility is always insured. The main recent modifications were done in the direction of being as much independent from the workflow-planner tool as possible. In this context, job script templates are now also stored in RefDB and provided as an item of the production instructions. Work is in progress in order to get these templates as general as possible to be usable by all sorts of farm setups (using Grid technologies or not, using mass storage system or not, ...). Work would also be required in order to implement a mechanism for checking compatibility between physics parameters, software versions, detector configuration, etc, in order to forbid invalid production requests As explained above, correspondence between executables and input/output file-types are recorded inside RefDB. In addition, some checks are being done within the scripts in very special cases. This issue is seen as being non-trivial if one wants to foresee all possibilities. The current solution is to trust that the physicists understand what they are asking for, and to offer them valid default options.

Finally, on physicists demand, prototype work is being done in order to provide the RefDB services to individual users, who want to profit from the book-keeping and production machinery for private data simulation outside the scope of large scale official productions and without connection to the central database.

## 7. CONCLUSIONS

The use, for large scale production, of RefDB, the CMS Central Reference Database and scripts, has shown to be very useful and has simplified the production machinery by allowing more automation of the system. It has increased the reliability of the data produced by reducing the amount of manual operations and improving validation. It has shown to be quite flexible: evolution of the system while keeping backward compatibility is not a difficult issue. It has been popular in the extent that isolated physicists have shown interest in having access to a similar tool for private production of data. Cloning of RefDB data and scripts has been prototyped by newcomers in the domain and the use of it has shown to be straightforward.

**Acknowledgments**

The authors want to thank Nick Sinanis, Werner Jank and Gilles Raymond for the installation and maintenance of the MySQL and web-servers used by RefDB.